\documentclass[12pt,letterpaper]{article}
\usepackage[font={sf,bf},width=6in]{caption}

\usepackage{epsfig,citesort,amssymb,amsmath,supertabular,subfigure,floatflt}

\usepackage[top=1.25cm,bottom=2.0cm,footskip=1.0cm,left=1.5cm, right=1.5cm,includeheadfoot,centering]{geometry} 

\usepackage[round]{natbib}
\usepackage{array}

\input{amssym.def}
\input{amssym}

\newcommand {\mm}[1] {\ifmmode{#1}\else{\mbox{\(#1\)}}\fi}

\newcommand{\utwi}[1]{\mbox{\boldmath $ #1$}}

\newcommand{\bx}{{\utwi{x}}}

\newcommand{\bD}{{\utwi{D}}}

\newcommand{\bI}{{\utwi{I}}}

\newcommand{\bP}{{\utwi{P}}}
\newcommand{\bQ}{{\utwi{Q}}}

\newcommand{\bS}{{\utwi{S}}}
\newcommand{\bT}{{\utwi{T}}}

\newcommand{\bV}{{\utwi{V}}}

\newcommand{\bLambda}{{\utwi{\Lambda}}}

\begin{document}

      \title{\bf
Estimation of amino acid residue substitution rates at local spatial
regions and application in protein function inference: A Bayesian
Monte Carlo approach }

      \author{\bf Yan Yuan Tseng and Jie Liang\thanks{Corresponding author. Phone:
      (312)355--1789, fax: (312)996--5921, email: {\tt
      jliang@uic.edu}} \\ Department of Bioengineering, SEO, MC-063 \\
      University of Illinois at Chicago\\ 851 S.\ Morgan Street, Room
      218 \\ Chicago, IL 60607--7052, U.S.A.}  \date{\today, In press, {\it Mol. Biol. Evol.  2006, 23: 421--436.}}

\maketitle
\vspace*{0.5in}

\vspace*{0.7in}
\noindent {\bf Running head:} Estimating Residue Rates by Bayesian Monte Carlo. 
\newpage
\abstract{\sf The amino acid sequences of proteins provide rich
information for inferring distant phylogenetic relationships and
for predicting protein functions.  Estimating the rate matrix of
residue substitutions from amino acid sequences is also important
because the rate matrix can be used to develop scoring matrices
for sequence alignment.  Here we use a continuous time Markov
process to model the substitution rates of residues and develop a
Bayesian Markov chain Monte Carlo method for rate estimation.  We
validate our method using simulated artificial protein sequences.
Because different local regions such as binding surfaces and the
protein interior core experience different selection pressures due
to functional or stability constraints, we use our method to
estimate the substitution rates of local regions.  Our results
show that the substitution rates are very different for residues
in the buried core and residues on the solvent exposed surfaces.
In addition, the rest of the proteins on the binding surfaces also have very
different substitution rates from residues.  Based
on these findings, we further develop a method for protein
function prediction by surface matching using scoring matrices
derived from estimated substitution rates for residues located on
the binding surfaces.  We show with examples that our method is
effective in identifying functionally related proteins that have
overall low sequence identity, a task known to be very
challenging.
}

\section*{Introduction}

Amino acid sequences are an important source of
information for inferring distant phylogenetic relationships and
for predicting the biochemical functions of protein.  Because the
substitutions of nucleotides can become rapidly saturated, and the
likelihood of unrelated identical substitutions is high for
nucleotides, the information of evolutionary conservation of
nucleotides is quickly obscured after a number of generations. The
mapping of DNA sequences by the genetic code to amino acid
sequences frequently can reveal more remote evolutionary
relation with more interpretable sequence similarity
\citep*{LioGoldman99_MBE}.  In addition, statistical analysis of
protein sequence alignment is also more reliable, as it is much
more difficult to detect and correct for deviations from
independent identical distributions in DNA sequences due to
possible translation of normal complexity DNA sequences into low
complexity protein sequences such as tandem repeats of simple
patterns of a few residues \citep*{Pearson_JMB98}.

The success in detecting evolutionarily related protein sequences
through sequence alignment depends on the use of a scoring matrix,
which determines the similarity between residues.  Rate
matrices of amino acid residue substitutions can be the basis for
the developing of many scoring matrices for sequence alignment.
Dayhoff {\it et al.\/} \citep*{pam} were the first to develop
empirical models of amino acid residue substitutions.  They used a
counting method to obtain accepted point mutation matrices (called
{\sc Pam} matrices).  The widely used {\sc Blosum} matrices can be
viewed as analogous to transition matrices of residues at
different time intervals \citep*{blosum,LioGoldman98_GR}.  They were
developed following a heuristic counting approach similar to that of
{\sc Pam}, and were derived from structure-based alignments of
blocks of sequences of related proteins \citep*{blosum}.  Both {\sc
Pam} and {\sc Blosum} matrices are widely used for sequence
alignment ({\it e.g.}, in software tools such as {\sc Fasta,
Blast} and {\sc Clustal W})
(\citealt{Altschul_JMB90}; \citealt*{Pearson_ME90,clustalw}).  An update of the
{\sc Pam} matrices based on the same counting approach using a
much enlarged database is the Jones-Taylor-Thornton ({\sc Jtt})
amino acid substitution matrix, which is widely used for
phylogenetic analysis \citep*{JTT,MOLPHY,Yang_CAB97}.

Whelan and Goldman pointed out that these counting methods are
effectively equivalent to the maximum parsimony method, and therefore
suffer from several drawbacks: the systematic underestimation
of substitutions in certain branches of a phylogeny and the
inefficiency in using all information contained in the amino acid
residue sequences \citep*{GoldmanWhelan_MBE}.  This can be a serious
problem for applications such as inferring protein functions from a
protein sequence, as the number of sequence homologs available for
multiple sequence alignment is often limited.  In addition, matrices
such as {\sc Pam} and {\sc Blosum} have implicit parameters whose
values were determined from the precomputed analysis of large quantities
of sequences, while the information of the particular protein of
interest has limited or no influence.  A more effective approach
for studying amino acid residue substitutions is to employ an explicit
continuous time Markov model based on a phylogenetic tree of the protein
\citep*{Yang_MBE98,GoldmanWhelan_MBE}. Markovian evolutionary models
are parametric models and do not have pre-specified parameter values.
These values are estimated from data specific to the protein of interest
\citep*{GoldmanWhelan_TG}.  Recent work using this approach has shown that
more informative rate matrices can be derived, with significant advantages
over matrices obtained from counting method \citep*{GoldmanWhelan_MBE}.

Despite these important results, current studies of the
substitution rates of amino acid residues are based on the
assumption that the whole protein sequence experience similar
selection pressure and therefore have the same substitution rates.
There is no distinction for different regions of proteins, namely,
all sites have the same evolutionary rates.  This is an
unrealistic assumption.  For example, regions that directly participate
in biochemical functions, such as binding surfaces, are likely to
experience very different selection pressure from other
regions.  In the protein interior, hydrophobic amino acid residues may be
conserved not due to their functional roles,
but due to the constraints of maintaining protein stability, as
hydrophobic interactions are the driving force of protein folding
\citep*{Dill_B90,Govindarajan_P97,Parisi_MBE01,LiLiang05_Proteins}.
Similarly, residues in the transmembrane segments of membrane proteins
experience different selection pressure from soluble parts of the proteins
\citep*{LioGoldman99_MBE,Tourasse_MBE00}.  It is therefore important to study
region-specific residue replacement rates.

An important advance in the reconstruction of phylogeny is the
consideration of heterogenous substitution rates among different sites
(\citealt{Yang_G00}; \citealt{Mayrose_MBE04}).  However, these are
based on substitution models of either nucleotides or codons, with
sometimes discretized categories of rates.  Because of the large number
of parameters due to an alphabet size of 20 for amino acid residues,
it is impractical to estimate site-specific rates for amino acid residue
sequences.

In this study, we use a continuous time Markov model to estimate
residue substitution rates for spatially defined regions of
proteins based on known three-dimensional structures of proteins
\citep*{Liang98_PS,Binkowski03_JMB}.  Different from previous
studies of rate estimation based on maximum likelihood methods
\citep*{Felsenstein_MBE96,Yang_MBE98,GoldmanWhelan_MBE,Siepel_MBE04},
we develop a Bayesian method to estimate the posterior mean values of the
instantaneous rates of residue substitution.  Our approach is based on the
technique of Markov chain Monte Carlo, a method that has been widely used
in phylogenetic analysis \citep*{Yang_MBE97,Mau_B99,Huelsenbeck_EIJOE00}.
To derive well defined spatial regions of proteins which are formed by
residues well separated in primary sequences, we rely on computational
analysis of protein structures \citep*{Liang98_PS}. In our study, these
distant residues in sequences are spatial neighbors that participate
in direct molecular binding events, and can be regarded as belonging to the
same class of substitution rates.  Our study is also motivated by the
need to deduce related functions from protein structures, {\it i.e.},
to identify functionally related protein structures.  As structural
biology proceeds, there is an increasing number of proteins whose atomic
structures are resolved, yet their biological functions are completely
unknown \citep{Andrzej}.

Our results show that residue substitution rates are significantly
different for different regions of the proteins, {\it e.g.}, for the
buried protein core, solvent exposed surfaces, and specific binding
surfaces on protein structures.  We also develop a novel method for
inferring protein functions.  Using residue similarity scoring matrices
derived from estimated substitution rates for protein surfaces, our
method is far more effective than several other methods in detecting
similar binding surface that are functionally related from different
protein structures.  This is a challenging task, as it is well known
that function prediction becomes difficult  when the sequence identity
between two proteins is below 60-70\% \citep*{Rost_JMB02,Tian_JMB03}.

This paper is organized as follows.  We first describe the
continuous time Markov model for residue substitution rates.  We
then discuss how to compute the likelihood function of
substitution rate matrices given a specific phylogeny and a multiple
sequence alignment.  The Markov chain Monte Carlo method is then
briefly described, including the design of move sets that helps to
improve the rate of mixing.  We then describe simulation results
in estimating substitution rates.  This is followed by discussion
of the results of different substitution rates estimated for
different regions of a  set of proteins. We then give examples to
show how residue scoring matrices derived from the estimated rate
matrix can improve detection of functionally related proteins.

\section*{Model and Methods}
\subsection*{Continuous time Markov process for residue
substitution.}  For a given phylogenetic tree, we use a reversible
continuous time Markov process as our evolutionary model
\citep*{Felsenstein81_JME,Yang94_JME}.
This model has several advantages over empirical methods.  For example,
Markovian evolutionary models are parametric models and do not have
pre-specified parameter values.  These values are all estimated from
data specific to the protein of interest \citep*{GoldmanWhelan_TG}.
In addition, previous works showed that the effects of secondary structure
and solvent accessibility are important for protein evolution, and such
effects can be captured by a Markovian evolutionary model, while it
is difficult for empirical methods to take these effects into account
(\citealt* {Jones96_JMB,Jones98_Genet,LioGoldman99_MBE}; \citealt{
Robinson_MBE03}).

Once the tree topology and the time intervals of sequence divergence
$\{t\}$ (or the branch lengths) of the phylogenetic tree are known,
the parameters of the model are the $20 \times 20$ rate matrix $\bQ$
for the 20 amino acid residues.  Because substitution rate and
divergence time $t$ are confounded, $t$ cannot be expressed in
absolute units. We follow the approach of \citep*{MOLPHY} to represent
the divergence time $t$ as the expected number of residue changes per
100 sites between the sequences.  The entries $q_{ij}$ of matrix $\bQ$
are substitution rates of amino acid residues for the set
$\mathcal{A}$ of 20 amino acid residues at an infinitesimally small
time interval.  Specifically, we have:
$\bQ = \{q_{ij}\}$, 
where the diagonal element is $q_{i,i} = -\sum_{i, j\ne i}
q_{i,j}$.
The transition probability matrix of size $20\times
20$ after time $t$ is
\citep*{LioGoldman98_GR}: $$
\bP(t) = \{ p_{ij}(t) \} = \bP(0) \exp (\bQ \cdot t), 
$$ where $\bP(0) = \bI$.  Here $p_{ij}(t)$ represents the probability
that a residue of type $i$ will mutate into a residue of type $j$
after time $t$.  To ensure that the nonsymmetric rate matrix $Q$ is
diagonalizable for easy computation of $\bP(t)$, we follow the
reference \citep*{GoldmanWhelan_MBE} and insist that $\bQ$ takes the
form of $\bQ = \bS \cdot \bD$, where $\bD$ is a diagonal matrix who
entries are the composition of residues from the region of interest on
the protein structure, and $\bS$ is a symmetric matrix whose entries
need to be estimated.  Because symmetric $\bS$ is diagonalizable as
$\bS = \bV \bLambda \bV^T$, the matrix $\bQ = \bS \cdot \bD =
\bD^{1/2}\bV \bLambda \bV^T \bD^{-1/2}$ is also diagonalizable,
hence
$\bP(t) 
= \bP(0) (\bD^{1/2} \bV) \exp (\bLambda t) (\bV^T \bD^{-1/2})$.

\subsection*{Likelihood function of a fixed phylogeny.}  
For node $k$ and node
$l$ separated by divergence time $t_{kl}$, 
the time reversible probability of observing
residue $x_k$ in a position $h$ at node $k$ and residue $x_l$ of the
same position at node $l$ is: $$
\pi_{x_k} p_{x_k x_l}(t_{kl}) = \pi_{x_l} p_{x_l x_k}(t_{kl}).
\label{reversible}$$
For a set $\mathcal{ S}$ of $s$ multiple-aligned sequences $(\bx_1,
\bx_2, \cdots, \bx_s)$ of length $n$ amino acid residues in a specific
region, we assume that a reasonably accurate phylogenetic tree $\bT =
(\mathcal{ V}, \mathcal{ E})$ of the proteins is given. Here $\mathcal
V$ is the set of nodes, namely, the union of the set of observed $s$
sequences $\mathcal L$ (leaf nodes), and the set of $s-1$ ancestral
sequences $\mathcal I$ (internal nodes). $\mathcal E$ is the set of
edges of the tree.  Let the vector $\bx_h = (x_1, \cdots, x_s)^T$ be
the observed residues at position $h$ for the $s$ sequences, $h$
ranges from $1$ to $n$.  Without loss of generality, we assume that
the root of the phylogenetic tree is an internal node $k$.  Given the
specified topology of the phylogenetic tree $\bf T$ and the set of
edges, 
the probability of
observing $s$ number of residues $\bx_h$ at position $h$ is:
\[
 p(\bx_h| \bT, \bQ) =
\pi_{x_k}
\sum_{\substack{
{i \in \mathcal{ I}}\\
{x_i \in \mathcal{ A}}
} }
\prod_{(i,j) \in \mathcal{ E}}
p_{x_i x_j}(t_{ij}).
\label{sameRate}
\]
after summing over the set $\mathcal{ A}$ of all possible residue types
for the internal nodes $\mathcal{ I}$.
The probability $P(\mathcal{ S}|\bT, \bQ )$
 of observing all
residues in the functional region is:
\[
P(\mathcal{ S}|\bT, \bQ ) = P(\bx_1, \cdots, \bx_s|\bT, \bQ) =\prod_{h=1}^n
p(\bx_h|\bT, \bQ).
\]
This can be used to calculate the log-likelihood function $
 \ell =  \log P(\mathcal{ S}|\bT, \bQ )$.

\subsection*{Bayesian estimation of instantaneous rates.}
Our goal is to estimate the values of the $\bQ$ matrix.  The continuous
time Markov model for residue substitutions has been implemented in
several studies using maximum likelihood estimation
\citep*{Yang94_JME,GoldmanWhelan_MBE}, and has also been applied in
a protein folding study \citep*{TsengLiang04_JMB}.  Different from these prior
studies, here we adopt a Bayesian approach.  We use a prior
distribution $\pi({\bQ})$ to encode our past knowledge of amino acid
substitution rates for proteins.
We describe the instantaneous substitution rate $\bQ = \{q_{ij}\}$ by a
posterior distribution $\pi(\bQ|\mathcal{ S}, \bT)$, which summarizes prior
information available on the rates $\bQ = \{q_{ij}\}$ and the information
contained in the observations $\mathcal{ S}$ and $\bT$.  After
integrating the prior information and the likelihood function, the posterior distribution
$\pi(\bQ|\mathcal{ S}, \bT)$ can be estimated up to a constant as:
\[
\pi(\bQ|\mathcal{ S},\bT) \propto \int
P(\mathcal{ S}|\bT, \bQ ) \cdot \pi(\bQ) d \bQ.
\]
Our goal is to  estimate
the posterior means of rates in $\bQ$ as  summarizing indice:
\[
\mathbb{E}_{\pi}(\bQ) = \int \bQ \cdot  \pi(\bQ|\mathcal{ S},\bT) d\bQ.
\]
In this study, we use uniform uninformative priors.  Others choices
are also possible.

\subsection*{Markov chain Monte Carlo.}
We run a Markov chain to generate samples drawn from the target
distribution $\pi(\bQ|\mathcal{ S},\bT)$.  Starting from a rate matrix
$\bQ_t$ at time $t$, we generate a new rate matrix $\bQ_{t+1}$ using the
proposal function: $
T(\bQ_t, \bQ_{t+1}).
$ The proposed new matrix $\bQ_{t+1}$ will be either accepted or
rejected, depending on the outcome of an acceptance rule
$r(\bQ_t,
\bQ_{t+1})$.  Equivalently, we have:
$$\bQ_{t+1} = A(\bQ_t, \bQ_{t+1}) =
T(\bQ_t, \bQ_{t+1}) \cdot r(\bQ_t,\bQ_{t+1}).$$
To ensure that the Markov chain will reach stationary state, we need to
satisfy the requirement of detailed balance, {\it i.e.},
$$
\pi(\bQ_t|\mathcal{ S},\bT)\cdot A(\bQ_t, \bQ_{t+1}) =
\pi(\bQ_{t+1}|\mathcal{ S},\bT)\cdot A(\bQ_{t+1}, \bQ_t).$$
This is achieved by using the Metropolis-Hastings acceptance ratio
$r(\bQ_t,\bQ_{t+1})$ to either accept or reject $\bQ_{t+1}$, depending
on whether the following inequality holds:
\[
u \le r(\bQ_t, \bQ_{t+1}) = \min \bigl\{ 1, \frac{
\pi(\bQ_{t+1}|\mathcal{ S},\bT) \cdot
T(\bQ_{t+1}, \bQ_t)}
{
\pi(\bQ_t|\mathcal{ S},\bT) \cdot
T(\bQ_t, \bQ_{t+1})} \bigr\},
\]
where $u$ is a random number drawn from the uniform distribution
$\mathcal{U}[0,1]$.  With the assumption that the underlying
Markov process is ergodic, irreducible, and aperiodic
\citep*{ProbBook}, a  Markov chain generated following these rules
will reach the stationary state \citep*{MCBook-Robert}.

We collect $m$ correlated samples of the $\bQ$ matrix after the Markov chain
has reached its stationary state. The posterior means of the rate
matrix are then estimated as:
\[
\mathbb{E}_\pi(\bQ) \approx \sum_{i=1}^m \bQ_i \cdot
\pi(\bQ_i|\mathcal{ S},\bT).
\]

\subsection*{Move set.}
A move set determines the proposal function
$T(\bQ_t, \bQ_{t+1})$, which is critical for the rapid convergency of a Markov
chain.  To improve mixing, we design two type of moves for proposing a
new rate matrix $\bQ_{t+1}$ from a previous matrix $\bQ_t$.  When the
state variable $s$ for these two types of moves takes the value $s=1$,
we take Type 1 move.  When the state $s=2$, we take Type 2 move.  For
Type 1 moves, a single entry of the rate matrix with index $ij$ is
randomly chosen, and with equal probability we assign: \[q_{ij, t+1} =
\alpha_1 q_{ij,t} \quad \mbox{ or } \quad q_{ij, t+1} = \alpha_2 q_{ij,t},\]
where $\alpha_1 = 0.9$, and $\alpha_2 = 1.1$.  For Type 2 moves, we
use a simplified residue alphabet of size 5 to represent the 20 amino
acid residue types, based on the analysis described in reference \citep*{LiLiang03_Proteins}.
The five residue types are: $\{ G, A, V, L, I, P\}$, $\{F,Y,W\}$,
$\{S,T,C,M,N,Q \}$, $\{D,E \}$, and $\{K, R, H \}$.
We select one of the 5 reduced residue types following $\mathcal{U}
[1,2, \cdots, 5]$, and scale with equal probability all entries
in $\bQ$ corresponding to the residues contained in one of the
simplified residue type, with a constant of either $\alpha_1=0.9$ or
$\alpha_2 = 1.1$ at equal probability.  The transition between these two
types of moves is determined by the transition matrix:
$$\begin{pmatrix} s_{1,1} & s_{1,2}\\ s_{2,1} & s_{2,2}\\
\end{pmatrix} =
\begin{pmatrix}
0.9 & 0.1\\
0.9 & 0.1
\end{pmatrix}.
$$
Overall, the acceptance ratio of Type 1 moves is $50\% - 66\%$, and
the acceptance ratio of Type 2 move is $<10\%$.

\subsection*{Rate matrix $\bQ$ and residue similarity score.}
To derive residue similarity scoring matrices for
 sequence alignments and database searches from the evolutionary model, we calculate the residue similarity scores
\citep*{Altschul_PNAS} $b_{ij}(t)$ between residues $i$ and $j$ at
different evolutionary time $t$ from the rate matrix $\bQ$:
\[b_{ij}(t)
= \frac{1}{\lambda} \log \frac{p_{ij}(t)}{\pi_j}
= \frac{1}{\lambda} \log \frac{m_{ij}(t)}{\pi_i \pi_j}
,
\] where $m_{ij}(t)$ is joint probability of
observing both residue type $i$ and $j$ at the two nodes separated by
time $t$, and $\lambda$ is a scalar.  Here $b_{ij}(t)$
satisfies the
equality $\sum \pi_i\pi _j e^{\lambda b_{ij}} =1$,
because of the
property of
the joint probability $\sum_{ij}m_{ij}(t)
= \sum_{ij} \pi_i p_{ij}(t) =  \sum_i \pi_i = 1$ holds for Markov matrix which has the property $\sum_j p_{ij}(t)$ \citep*{ProbBook}.
The overall expected score of this matrix is then $\sum_{ij}m_{ij}(t)b_{ij}(t)$, usually in bit units \citep*{Altschul_PNAS}.

\subsection*{Computation of surface pockets and interior voids.}

We use the {\sc Volbl} method to compute the solvent accessible
surface area of protein structures (\citealt{Edels95_Hawaii,Liang_P98}).  We
use the {\sc CastP} method \citep*{Liang98_PS,castp_NAR} to identify
residues located on surface pockets.  Both {\sc Volbl} and {\sc CastP}
are based on precomputed alpha shapes \citep*{Edels94_ACMTG},
where the dual simplicial complex is constructed from the Delaunay
triangulation of the atomic coordinates of the protein.  We use the
pocket algorithm \citep*{Edels98_ADM,Liang98_PS} in {\sc CastP} to identify
residues located in surface pockets and interior voids.  Details and other
applications of these methods can be found in
\citep*{Liang98_PS,Edels98_ADM,Liang_BJ01,Binkowski03_JMB}.

\section*{Results}
There is a large number of parameters (189) characterizing the
substitutions of amino acid residues.  We first need to understand
at what accuracy these parameters can be estimated.  Because we
are studying regions ({\it e.g.}, binding surfaces) on a protein
structure, we often only have a few dozen instead of a few hundred
residue positions available for parameter estimation. In addition,
we are frequently limited by the available sequence data, and
the size of the phylogenetic tree may be moderate.  Even if the
parameters of the substitution model can be estimated, it is not
clear how effective they are for applications such as inferring
protein functions from protein structures.  We describe our
results addressing each of these issues.

\subsection*{Rate estimation: simulation studies.}
We first carry out a simulation study to test
the accuracy of the estimated residue substitution rates.  We generate
a set of artificial sequences based on an evolutionary model with
known substitution rates.  We ask whether our method can
recover the original substitution rates reasonably well, and how
many sequences and residues are necessary so an accurate estimation can
be made.  For this purpose, we first take the sequence of the
alpha-catalytic subunit of cAMP-dependent protein kinase (SwissProt
P36887, pdb {\tt 1cdk}, with length 343), and the sequence of
carboxypeptidase A2 precursor (SwissProt P48052, pdb {\tt 1aye},
length 417).

\paragraph{Statistics for estimation accuracy.}
We use the Jones-Taylor-Thornton ({\sc Jtt}) evolutionary model
\citep*{JTT}, which is characterized by a frequency-independent amino
acid interconversion rate matrix $\bS_{\mbox{JTT}}$ and the diagonal
matrix $\bD$ of the composition of the 20 amino acid residues for the
set of sequences that were used to derive the original {\sc Jtt} model
\citep*{Yang_CAB97}.  The substitution rate matrix $\bQ_{JTT}$ is
then: $\bQ_{JTT} = \bS_{JTT} \cdot \bD$.  To avoid potential bias, we
use the composition $\bD$ of the protein kinase and the
frequency-independent amino acid interconversion rate matrix of
$\bS_{JTT}$ to calculate the instantaneous rate matrix $\bQ$ for
the protein kinase, which is then used to generate 16 artificial kinase
sequences at different time intervals $t$ using the probability $\bP(t)
= \exp (\bQ t) \bI$.  Here we use a simple balanced phylogenetic tree
of 16 leaf nodes with equal branch lengths of $t=0.1$ for all edges.
We compare the estimated frequency-independent amino acid
interconversion rate matrix $\widetilde \bS$ to the true matrix
$\bS_{JTT}$.

For comparison, we first normalized the estimated and true
{\sc Jtt} frequency-independent interconversion rate matrices, such
that:
$$
\frac{1}{20}
\sum_{ij, \; i\ne j} s_{ij} = 1 \quad \text{and}
\quad
\frac{1}{20} \sum_{ij, \; i\ne j} \tilde{s}_{ij} = 1,
$$
where $s_{ij}$ is the $(i,j)$-th entry of the matrix $\bS$.

We are interested in the rates of substitution that occur in a
specific spatial region of the protein.  Because these regions
contain only a subset of the residues and often are under
different selection pressure, not all possible substitutions are
observed with adequate frequency for estimation.  In addition, the
usually moderate size of the phylogenetic tree limits the observed
frequency of some substitutions.  Nevertheless, the frequently
observed substitutions for a specific protein region are likely to
be the most important ones, and the estimation of their rates
should be better than rates of infrequently observed
substitutions.

We need to quantitatively assess our estimation error.
Because it is very difficult to estimate accurately the absolute
values of the individual rates, we assess instead the errors in
estimated $\tilde{s}_{ij}$ in terms of their effects on the overall patterns
of residue substitution on a specific protein region.  This is more
appropriate for many applications such as the analysis of the
evolution of binding surfaces and the evolution of the folding core, as only
a subset of substitutions occur at a functional surface or in the core.  We develop
some quantitative measures for this purpose.

We call a residue pair $(i,j)$ an {\it occurring pair\/} if both
residues $i$ and $j$ occur simultaneously in one column of the
multiple aligned sequences of a specific region.  For the subset
of rates ${\cal S} = \{s_{ij}\}$ for a residue pair  $(i,j)$  from
the set of occurring pairs $\cal P$, we obtain the {\it relative
contribution\/} of a specific frequency-independent
interconversion rate between a pair of residues as:
$$ s'_{ij} = s_{ij}/\sum_{ij
\in {\cal P}} s_{ij}.
$$

The $\Delta e_{ij}$
{\it weighted error in contribution\/} is computed as:
$$
\Delta e_{ij} \equiv \frac{f_{ij}}{\sum_{ij,  i \ne j}f_{ij}
}[\tilde{s}'_{ij} - s'_{ij}],
$$
where $\tilde{s}_{ij}$ is the estimated value of $s'_{ij}$,
$f_{ij}$ is the number count of how often the $(i,j)$
substitutions occur.

To measure the overall differences of the estimated $\tilde{\bS}$
and the original $\bS_{\mbox{JTT}}$ matrices for the occurring
substitutions, we use the {\it weighted mean square error\/}
($MSE_{\cal P}$) \citep{Mayrose_MBE04}:
$$
MSE_{\cal P} \equiv \sum_{ij \in {\cal P}} \Delta e_{ij}^2/|{\cal P}|.
$$

\paragraph{Error analysis in estimated rates.}

\begin{figure}[!t]
\begin{center}
\epsfxsize=4in %width of figure - will enlarge/reduce the figures
\epsfbox{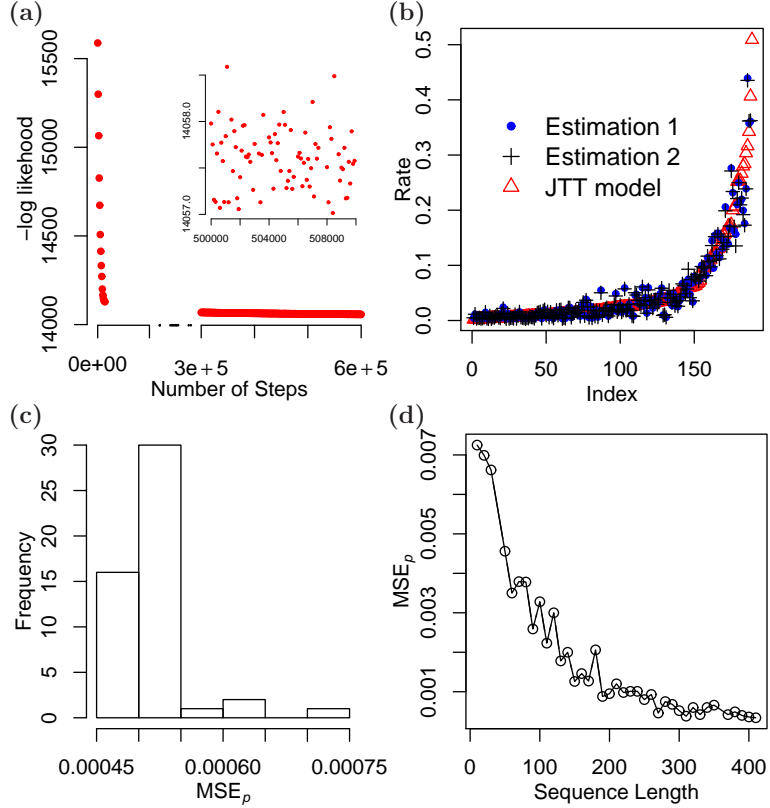}
\caption{
{\small \sf Estimating residue
substitution rates using simulated carboxypeptidase sequences. (a) The
Markov chain converges after $3\times10^{5}$.  The insert shows negative
log likelihood ($-\ell$) values in stationary state after the burning-in
period; (b) $s_{ij}$ values estimated in two simulations are all similar
to the true rates.  In the first simulation, the 189 initial values are
set such that $s_{i,j} = 0.1$ for all entries.  In the second simulation,
the 189 initial $s_{i,j}$ values are sorted numerically by index $i$
then by index $j$, and the values are assigned from $0.1$ with an
increment of $0.01$ for the next entry. (c) The $MSE_{\cal P}$ values
from 50 repeated estimations of substitution rates of carboxypeptidase
with random initial values are all less than $8 \times 10^{-4}$. The
mean value of $MSE_{\cal P}$ is $5.2 \times 10^{-4}$. (d) The value
of $MSE_{\cal P}$ depends on the length of available subsequences.
For subsequence of length $\geq 20$, the $MSE_{\cal P}$ value is $<0.008$. }
}
\label{fig:1}
\vspace *{-10 pt}
\end{center}
\end{figure}

Using the 16 artificial sequences generated from the sequence of
carboxypeptidase and a simple balanced phylogenetic tree with equal
branch length $t=0.1$ for all edges between nodes, 
the Markov chain
converges quickly after $3\times 10^5$ Monte Carlo steps
(Figure~\ref{fig:1}a), as shown by the value of $-\ell$ for the negative
likelihood function.  After a burning-in
period of $3 \times 10^5$ steps, we collect $m = 4\times 10^5$ samples
for estimating $\{ s_{ij} \}$ values.  Figure~\ref{fig:1}b shows the
estimation results for two simulations started from two different sets
of initial values of $\{s_{ij} \}$.  It is clear that both sets of
estimated rates $\{ \tilde{s}_{ij} \}$ for the occurring pairs are in
general agreement to the set of true values from the  {\sc Jtt} model.

To further assess how robust the estimations are, we repeated the
Markov chain Monte Carlo simulation 50 times using random initial
values of $\{s_{ij} \}$ drawn from a uniform distribution of
${\cal U}(0,1)$.  On average, the estimation error is small. The
mean of the overall weighted $MSE_{\cal P}$ from 50 simulations is $5.2 \times
10^{-4}$ for occurring pairs (Figure~\ref{fig:1}c).

\paragraph{Length dependency of errors in estimated parameters.}

To estimate region specific substitution rates, it is important to
assess how the accuracy of the estimation depends on the size of the
region.  For example, the functional region of a protein contains
only a small number of residues, which varies depending on the
size of the binding site.  We carry out another simulation study
for this purpose.  Starting from the N-termini of the 16
artificially generated carboxypeptidase sequences, we take a
subsequence from each sequence, with the length increasing from 10
to 417, at an increment of 10 residues.  We then estimate the
substitution rates at each length.  Each simulation of a different
length was started from a random set of initial values drawn from
${\cal U}(0, 1)$, and the same burning-in period and sample size $m$
are used as before.  The $MSE_{\cal P}$ values obtained using sequences of
different lengths are plotted in Figure~\ref{fig:1}d.  Our results
show that for this set of sequences, as long as the number of
residues is $\ge 20$, the $MSE_{\cal P}$ of the estimated parameters will
be less than $0.008$.

\begin{figure}[!t]
\begin{center}
\epsfxsize=4in   %width of figure - will enlarge/reduce the figures
\epsfbox{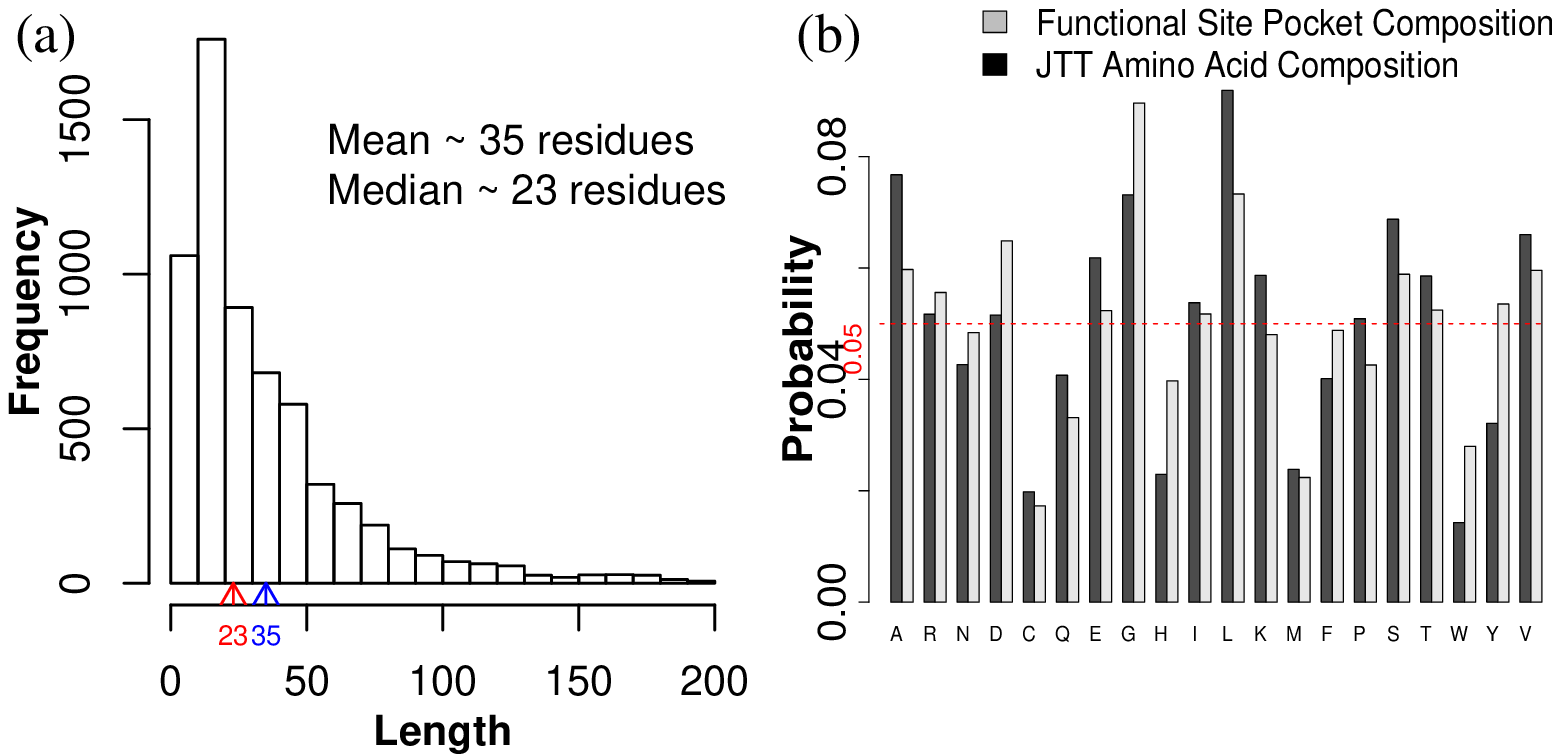}
\caption{{\small \sf The length
distribution and amino acid composition of functional pockets.
(a) The length distribution of $6,273$ functional pockets. The
average length of functional pockets is 35 residues, and the
median is 23 residues.  (b) Comparison of amino acid compositions
of residues in $6,273$ functional pockets with the composition of
$16,300$ protein sequences used to derive the {\sc Jtt}
substitution matrix.  The dashed line is the expected probability
of $0.05$ if all substitution rates following the uniform
distribution. }}
\label{fig:PocDist}
\vspace *{-10 pt}
\end{center}
\end{figure}

Based on analysis of the protein structures in the Protein Data
Bank, we found that among the surface pockets from 6,273 protein
structures that all contain annotated functional residues (as recorded
either in the {\sc Feature} table of the {\sc SwissProt} database or
the {\sc Active Site} field of the {\sc Pdb} file), the average size
of a functional site pocket is 35 residues, and the median is 23
residues (Figure~\ref{fig:PocDist}a).  This suggests that our method will
be applicable for the analysis of protein functional pockets.

\begin{figure}[!t]
\begin{center}
\epsfxsize=4in   %width of figure - will enlarge/reduce the figures
\epsfbox{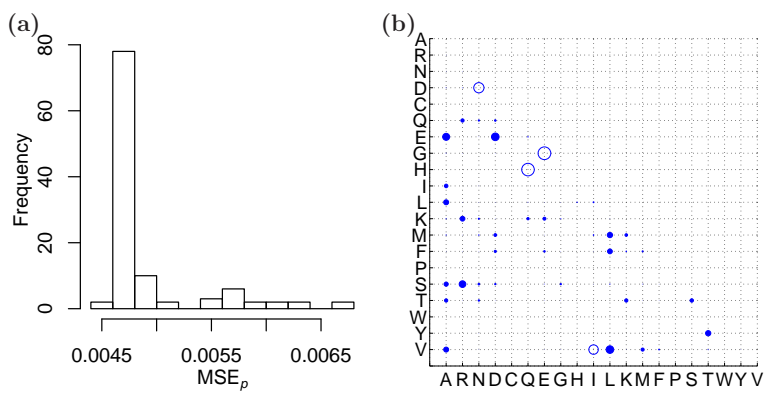}
\caption{{\small \sf Estimating the
substitution rates of residues on the binding surface of
cAMP-dependent protein kinase from simulated sequences.  (a) For
110 independent estimations of the substitution rates with random
initial values, the $MSE_{\cal P}$ values are all $<8 \times 10^{-4}$.  The
mean $MSE_{\cal P}$ value of the 110 estimations is $0.0048$. (b) There are
only 4 substitutions (empty circles) whose error $\Delta e_{ij}$
is great than $3.0\%$, although all of the 90 occurring pairs have
$\Delta e_{ij}<4.5\%$. }}
\label{fig:3}
\vspace *{-10 pt}
\end{center}
\end{figure}

We carried out another simulation study estimating substitution
rates only for the binding surface of a protein.  Using the same
phylogenetic tree as that of the carboxypeptidase simulations and
the same {\sc Jtt} model, we generate 16 artificial sequences of
the alpha-catalytic subunit of cAMP-dependent protein kinase
(SwissProt P36887, pdb {\tt 1cdk}, length 343).  Our goal is to
estimate rates only for the subset of 38 residues located in the
binding site.  Figure~\ref{fig:3}a shows that the $MSE_{\cal P}$ values of
the estimated rates from 110 independent simulations for the 90
occurring pairs of residues are all small.  The estimated rates
from all simulations have $MSE_{\cal P}$ $< 8 \times 10^{-3}$, and the mean
of the overall $MSE_{\cal P}$ from 110 simulations is $4.8 \times 10^{-3}$ for
the 90 occurring pairs.  Clearly, the estimation errors measured
in $MSE_{\cal P}$ are larger when only residues in the binding site are used
compared to the estimation errors of carboxypeptidase where all $417$
residues are used.  Nevertheless, the estimations are still
useful, as the mean $MSE_{\cal P}$ value remains small. Figure~\ref{fig:3}b
plots the individual mean value of weighted errors $\Delta e_{ij}$
for the 90 occurring pairs obtained from 110 simulations. There
are only 4 substitutions whose weighted error in contribution
$\Delta e_{ij}$ is greater than $3\%$ , although all occurring
pairs have $\Delta e_{ij}<4.5\%$.

\subsection*{Evolutionary rates  are region specific.}

\paragraph{Exposed surface and buried interior have different substitution rates.}

Residues on protein surfaces that are exposed to solvent are under
different physicochemical constraints from residues in the buried
interior.  We estimate the substitution rates for exposed and
buried regions on a protein structure.  We use a simple criterion to classify residues as either
exposed or buried: Based on the calculation of solvent accessible (SA)
surface area using {\sc Volbl} \citep*{Liang_P98}, we declare a residue to be buried if
its SA area is 0 $\rm{\AA}^2$, and exposed if SA area $> 0$ $\rm{\AA}^2$.

\begin{figure}[!t]
\begin{center}
\epsfxsize=4in   %width of figure - will enlarge/reduce the figures
\epsfbox{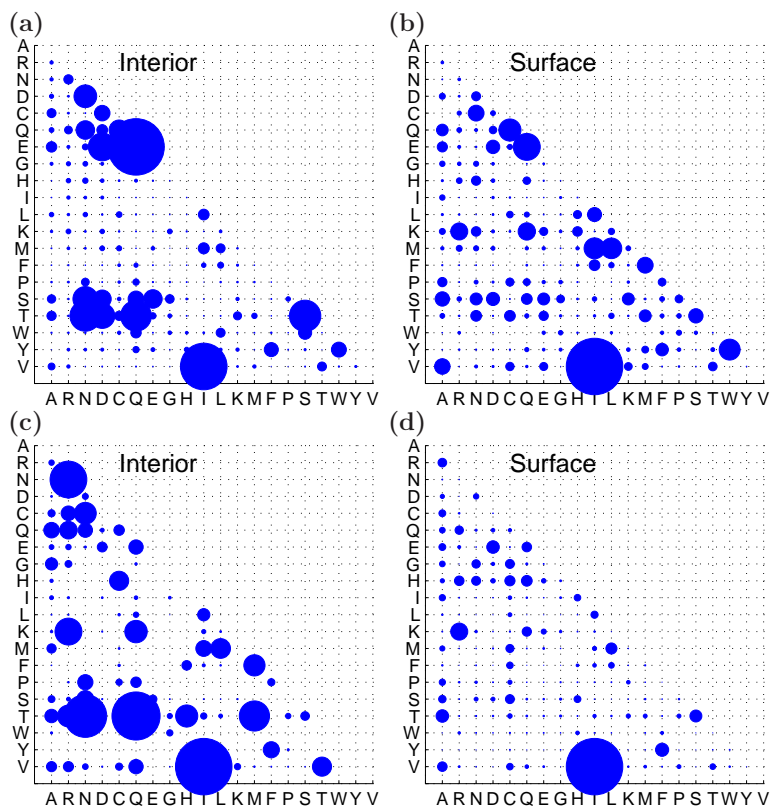}
\caption{{\small \sf Substitution rates
of residues on solvent exposed surface and in buried interior. (a)
Substitution rates of buried interior residue on 2-haloacid
dehalogenase (pdb {\tt 1qh9}).  There are 100 occurring pairs. (b)
substitution rates of surface exposed residues of {\tt 1qh9}.
There are 188 occurring pairs. (c) Substitution rates of buried
interior residue of alpha amylase (pdb {\tt 1bag}). There are 190
occurring pairs. (d) substitution rates of surface exposed
residues of {\tt 1bag}.  There are 177 occurring pairs. }}
\label{fig:bubble}
\vspace *{-10 pt}
\end{center}
\end{figure}

For the protein 2-haloacid dehalogenase (pdb {\tt 1qh9}),
Figure~\ref{fig:bubble} shows that the residues on the exposed surfaces
and in the buried interior have very different substitution patterns.
For example, the substitution of Threonine (T) with Asparagine (N),
Aspartate (D), or Glutamine(Q) occurs much more frequently in the
buried interior than on the surface (Figure~\ref{fig:bubble}a
and Figure~\ref{fig:bubble}b).  A similar pattern is also seen
for alpha amylase (pdb {\tt 1bag}, Figure~\ref{fig:bubble}c and
Figure~\ref{fig:bubble}d).  In general, ionizable and polar residues in
the protein interior have higher propensities to mutate to other ionizable
and polar residues.

The frequent substitutions between T and $\{\mbox{N, D, Q}\}$
observed in the protein interior of l-2-haloacid dehalogenase and
amylase suggest that to maintain the H-bonding interactions in
the protein interior, it is far more common to have substitutions
among ionizable residues and polar residues.  These substitution
patterns point to the importance of preserving polar interactions,
which provide important structural stability in the protein interior,
as the high dielectric constants inside proteins makes the
electrostatic contribution of salt-bridges and H-bonds in the protein
interior stronger than H-bonds on protein surfaces.

\begin{table}
\captionsetup{labelsep=newline,singlelinecheck=false}
\begin{center}
\caption{
\bf Substitutions rate of residues in the interior and on the exposed surface are different.}
\vspace*{10pt}
\label{tab:ISrate}
\begin{tabular}{lcccc}
\hline
\hline
Protein Family&pdb&Interior&Surface&$p-$value of K-S test
\\
{}&{}&occurring pairs& occurring pairs&{}\\
\hline
EC 3.4.11.18&{\tt 1b6a}&80&175&0.016\\
EC 3.2.1.1&{\tt 1bag}&190&177&0.015\\
EC 2.3.3.1&{\tt 1csc}&55&163&0.009\\
EC 3.8.1.2&{\tt 1qh9}&139&169&0.023\\
EC 3.2.1.21&{\tt 1h49}&60&169&0.024\\
EC 3.5.1.5&{\tt 1udp}&92&162&0.014\\
EC 1.1.1.37&{\tt 1b8v}&97&150&$4.8 \times 10^{-5}$\\
\hline \hline
\end{tabular}
\end{center}
\end{table}

The conclusion that residues in the protein interior experience different
selection pressure from residues on the protein surfaces are likely to be true
for other proteins.  We estimated the substitution rates of buried residues
and exposed residues for 6 additional proteins with different
biological functions as indicated by different enzyme classification
numbers (Table~\ref{tab:ISrate}).  In all cases, we find that surface
residues have different evolutionary patterns overall.  Although not all
substitution rates are noticeably different, Table~\ref{tab:ISrate} shows that
for each of the 8 proteins studied, we can reject the null hypothesis, based
on the nonparametric Kolmogorov-Smirnov test, that the two distributions
of substitution rates for the set of exposed residues and the set of
buried residues are the same.

\paragraph{Residues in functional sites and on the rest of the surface
have different substitution rates.}

\begin{figure}[!t]
\centerline{
\epsfxsize=6in   %width of figure - will enlarge/reduce the figures
\epsfbox{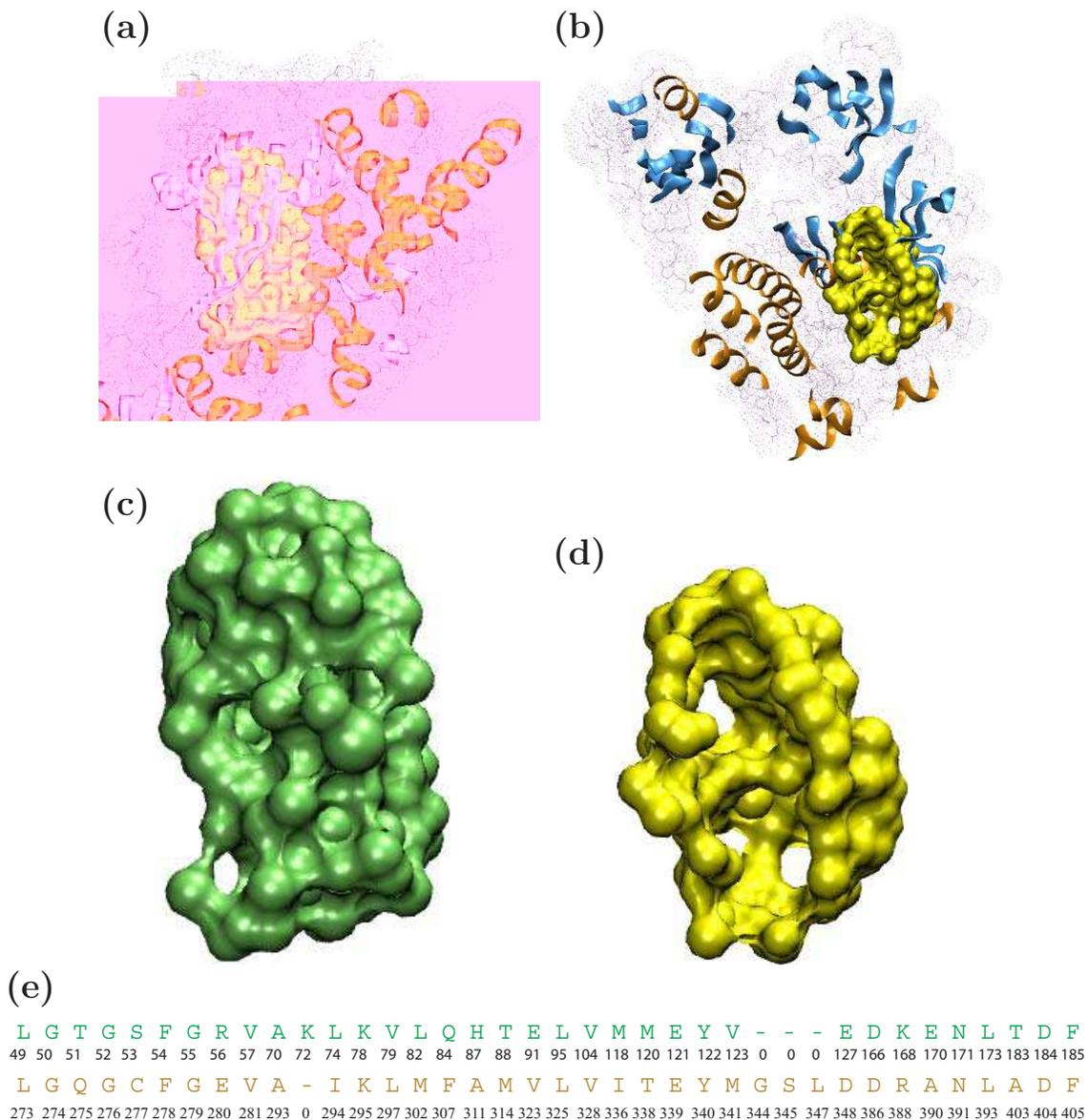}}
\caption{{\small \sf Protein functional
pockets of kinases.  Functional site of (a) the catalytic subunit
of cAMP-dependent protein kinase ({\tt 1cdk} chain A), and (b)
tyrosine protein kinase c-src ({\tt 2src}). Both kinases bind to
AMP or AMP analogs. Their global primary sequence identity is as
low as 16\%.  However, if we extract their binding surfaces (as
shown in (c) and (d)) out, (e) the residues forming the binding
pockets have much a higher sequence identity (51\%). }}
\label{fig:kinase}
\vspace *{-10 pt}
\end{figure}

Protein functional sites are the regions where a protein interacts with
ligand, substrate, or other molecules.  Because proteins fold into
their three-dimensional native structures, functional sites often
involve residues that are distant in sequence but are in spatial
proximity.  As can be seen in Figure~\ref{fig:kinase}, two proteins with a low sequence identity ($<16\%$)  may
be very different overall, but their
functional binding pockets may be quite similar.  In this study, we use
the {\sc CastP} database of precomputed surface pockets for our
analysis of functional sites on protein structures.  This approach has
been applied in studies of protein function prediction
\citep*{Binkowski03_JMB,castp_NAR} and in structural analysis of non-synonymous
SNPs \citep{Stitziel_JMB03}.

Residues that are located in functional pockets are under
different selection pressures.  This can be clearly seen in
Figure~\ref{fig:PocDist}b, such that the composition of residues in 
functional pockets is very different from the composition of residues in the set of
full protein sequences from which the {\sc Jtt} substitution
matrix was derived.  Here we examine only protein surface pockets
that contains functionally important residues as annotated by
either {\sc SwissProt} or {\sc Pdb}.  In functional pockets, Tyr,
Trp, His, Asp and Gly residues are far more enriched, but Leu,
Ser, and Ala are less if compared to sequences used in the {\sc Jtt} rate
matrix analysis.  Tyr, Trp, His and Asp are residues that play
important roles in enzyme reactions through electrostatic
interactions, change of protonation states, and aromatic
interactions.  Gly residues are important in the formation of turns and other
geometric features for binding site formation.  The enrichment of
hydrophobic Leu and small residues Ser and Ala in the full
sequence are probably important for structural stability.

\begin{figure}[!t]
\begin{center}
\epsfxsize=4in   %width of figure - will enlarge/reduce the figures
\epsfbox{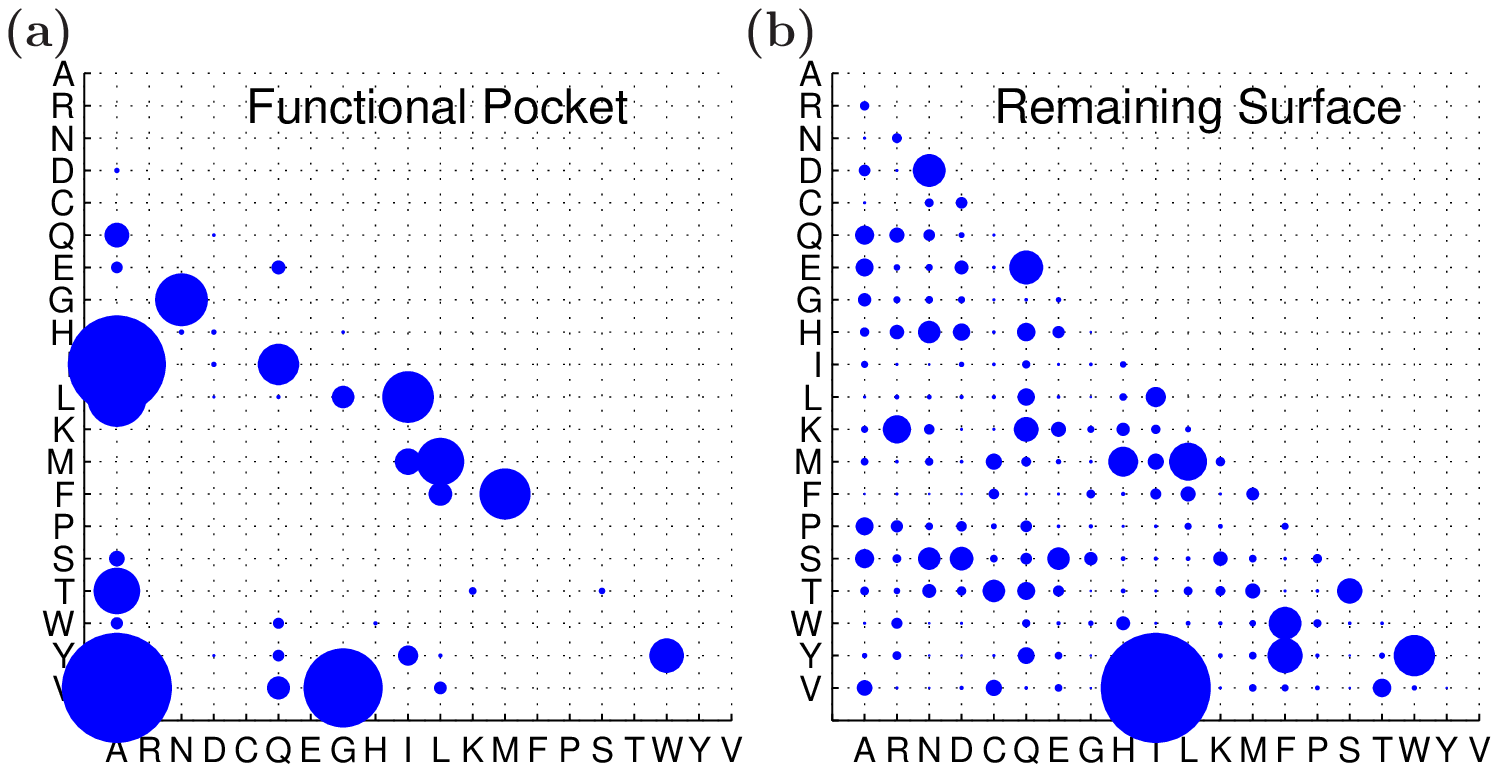}
\caption{{\small \sf Substitution
rates of residues in the functional binding surface and the
remaining surface of alpha-amylase (pdb {\tt 1bag}). (a)
Substitution rates of the functional binding surface.  There are 39
occurring pairs. (b) substitution rates of the remaining surface
on {\tt 1bag}. There are 177 occurring pairs. }}
\label{fig:amylaseBubble}
\vspace *{-10 pt}
\end{center}
\end{figure}

We examine the patterns of residue substitutions on protein functional
surfaces in some detail.  Taking a structure of alpha amylase (pdb
{\tt 1bag}) as an example, we compare the estimated substitution rate
matrix of functional surface residues with that of the remaining
surface residues of the protein (Figure~\ref{fig:amylaseBubble}).
It is clear that the selection pressures for residues located in
functional site and for residues on the rest of the protein surface
are different, and they are also both
different from the {\sc Jtt} matrix (data not shown).  This suggests that identifying
functionally related protein surfaces will be more  effective if
we employ scoring matrices specifically derived from residues located on
functional surface instead of using a general precomputed substitution
matrix.

\subsection*{Application: Detecting functionally similar biochemical binding surfaces.}
For proteins carrying out similar functions such as binding similar
substrates and catalyzing similar chemical reactions, the binding
surfaces experience similar physical and chemical constraints.  The
sets of allowed and forbidden substitutions will
therefore be similar because of these constraints.  The continuous time
Markov model can provide evolutionary information at different time
intervals once the instantaneous substitution rates are estimated.
This information is encoded in the time-dependent residue
substitution probabilities.  An objective test of the utility of the
estimated evolutionary model is to examine if we can discover functionally related
proteins, namely, whether we can identify protein
structures that have similar binding surfaces and carry out similar
biological functions.

\paragraph{Identification of functionally related proteins  from a
template binding surface.}

We use alpha-amylases as our test system.  Alpha-amylase (Enzyme
Classification number E.C.3.2.1.1) acts on starch, glycogen and related
polysaccharides and oligosaccharides.  Detecting functionally related
alpha amylase is a challenging task, as many of them have very
low overall sequence identities ($<25\%$) to the query protein template.  If two
proteins have a sequence identity below $60-70\%$, it becomes difficult to
make functional inferences based on sequence alignment \citep*{Rost_JMB02}.

Given a template binding surface from an alpha amylase ({\tt
1bag}, pdb), we wish to know how many protein structures can be
identified that have the same enzyme classification (E.C.)\
number at an accuracy of all four E.C.\ digits.  These protein
structures all carry out the same or related reactions.  By the
convention of the Enzyme Classification system, the E.C.\ numbers
represent a progressively finer classification of the enzyme, with
the first digit about the basic reaction, and the last digit often
about the specific functional group that is cleaved during
reaction.

\begin{figure}[!t]
\centerline{
\epsfxsize=4.in   %width of figure - will enlarge/reduce the figures
\epsfbox{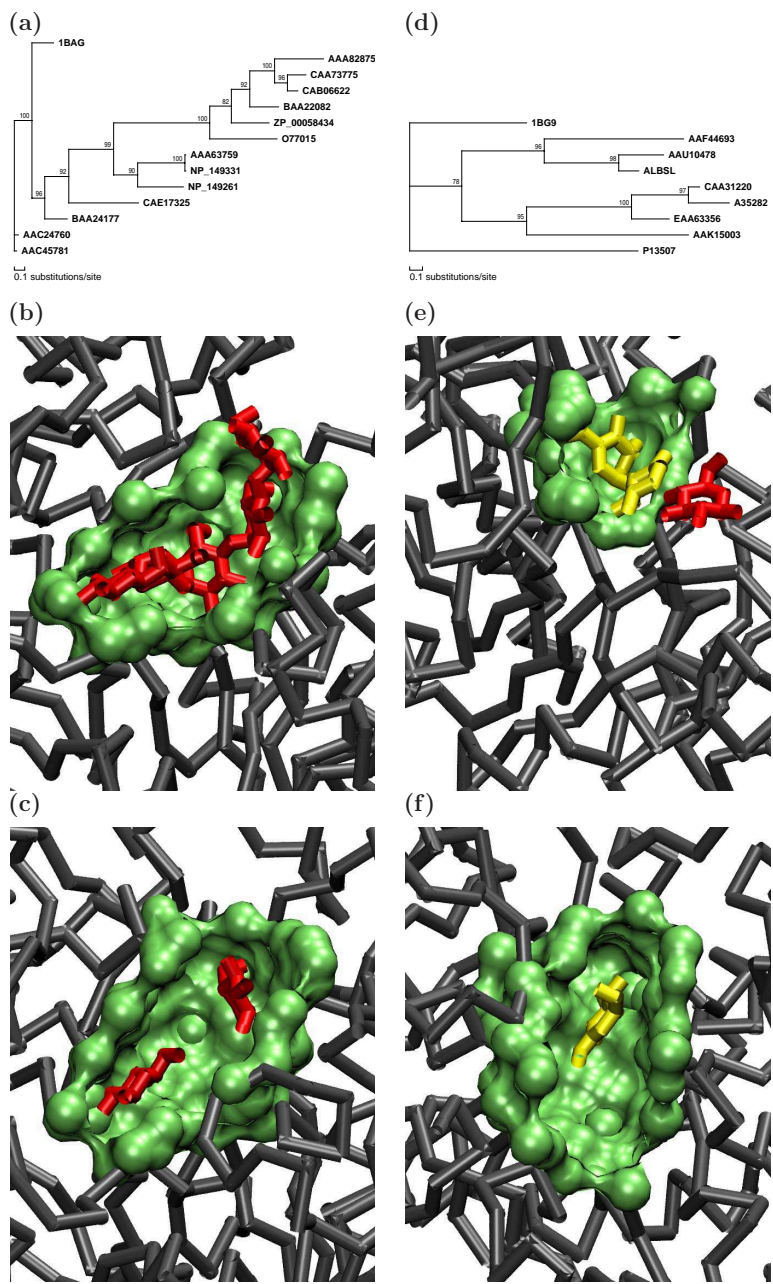}}
\caption{{\small \sf Function prediction of alpha amylases.  (a) The
phylogenetic tree for {\sc Pdb} structure {\tt 1bag} from {\it B.\
subtilis}.  (b) The functional binding pocket of alpha amylase on {\tt
1bag}.  (c) A matched binding surface on a different protein structure
({\tt 1b2y} from human, full sequence identity 22\%) obtained by
querying with the binding surface of {\tt 1bag}.  (d) The phylogenetic
tree for {\tt 1bg9} from {\it H.\ vulgare}.  (e) The binding pocket on
{\tt 1bg9}.  (f) A matched binding surface on a different protein
structure ({\tt 1u2y} from human, full sequence identity 23\%)
obtained by querying with {\tt 1bg9}. }}
\label{fig:2}
\vspace *{-5 pt}
\end{figure}

We first exhaustively compute all of the voids and pockets on this protein
structure \citep*{Liang98_PS,castp_NAR}.  Based on biological annotation
contained in the Protein Data Bank, the 60th pocket containing 18
residues is identified as the functional site (Figure~\ref{fig:2}b).
To construct an evolutionary model, we use sequence alignment tools
to gather sequences homologous to that of {\tt 1bag}
\citep{Altschul_NAR97}.  After removing redundant sequences and
sequences with $>90\%$ identity to any other identified sequences or
the query sequence of {\tt 1bag}, we obtain a set of 14 sequences of
amylases.  These 14 sequences are used to construct a phylogenetic
tree of alpha-amylase (Figure~\ref{fig:2}a).  We use the
maximum-likelihood method implemented in the {\sc Molphy} package
for tree construction \citep*{MOLPHY}.

We then calculate the similarity scoring matrices from the estimated
values of the rate matrix.  Because {\it a priori\/} we do not know
how far a particular candidate protein is separated in evolution from
the query template protein, we calculate a series of $300$ scoring
matrices, each characterizing the residue substitution pattern at a
different time separation, ranging from 1 time unit to 300 time unit.
Here 1 time unit represents the time required for 1 substitution per
100 residues \citep*{pam}.  We use the Smith-Waterman algorithm as
implemented in the {\sc Ssearch} method of {\sc Fasta}
\citep*{Pearson_G91} with each of the $300$ scoring matrices in turn
to align sequence patterns of candidate binding surfaces from a
database of $>$2 million protein surface pockets contained in the {\sc
pvSoar} database \citep*{pvSOAR}.  We use an $E$-value
of $10^{-1}$ as the threshold to decide if a matched surface pocket is a hit.  Surfaces
similar to the query binding pocket identified (with $E$-values
$<10^{-1}$) are then subjected to further shape analysis, where those
that cannot be superimposed to the residues of the query surface
pattern at a statistically significant level ($p$-value $<0.01$) by
either the coordinate RMSD measure or the orientational RMSD
\citep*{Binkowski03_JMB}measure are excluded.  The $p$-value is
estimated using methods developed in \citep*{Binkowski03_JMB}.

A total of $58$ PDB structures are found to have similar binding
surfaces to that of {\tt 1bag}, and hence are predicted as amylases.
All of them turn out to have the same E.C.\ number of 3.2.1.1 as that of {\tt
1bag}.  We repeat this study but using a different amylase structure
as the query protein.  Using the functional pocket on {\tt 1bg9}, we
found $48$ PDB structures with  E.C.\ 3.2.1.1 labels.  The union of the
results from these two searches gives 69 PDB structures with
E.C.3.2.1.1 labels.  Examples of matched protein surfaces are shown in
Figure~\ref{fig:2}.

\paragraph{Comparison with others.}
We compare our results with other studies.  The Enzyme Structure
Database ({\sc Esd}) ({\tt www.ebi.ac.uk/thornton-srv}) collects
protein structures for enzymes contained in the {\sc Enzyme} databank
\citep*{Bairoch_NAR93} for study.  Here we take the ESD database as
the gold-standard, and all true answers are contained in this human
curated database.  There are $75$ PDB entries with enzyme class label
E.C.3.2.1.1 in {\sc Esd} (version Oct 2004).  Out of the $75$
structures, our method discovered $69$ PDB structures (no redundancy)
using {\tt 1bag} and {\tt 1bg9} as queries.

We also compare our results with those obtained from a database search
using sequence alignment methods.  Using the Smith-Waterman algorithm
as implemented in {\sc Ssearch} of the {\sc Fasta} package with the
default {\sc Blosum50} matrix, only 32 structures are identified as
alpha amylase (see Table~2 in \citep*{Binkowski03_JMB}).  When using
{\sc Psi-blast} and the NR database with default parameters, an
$E$-value threshold of $10^{-3}$, and $<10$ iterations to generate
position-specific weight matrices, $65$ structures (no redundancy) among
the $75$ known structures of alpha-amylase are found after combining
results from queries with {\tt 1bag} and {\tt 1bg9}.

We next tested search results using the standard {\sc Jtt} matrix instead
of the estimated protein-specific and surface-specific matrix. In this
case, we find 52 hits instead of 58 using {\tt 1bag} as the query protein,
and 8 hits instead of 48 using {\tt 1bg9} as the query protein.

Our method differs from {\sc Ssearch} \citep*{Pearson_JMB98} in two
aspects: first, we use short sequence patterns generated from the
binding surface of the protein structure instead of the full protein
sequences.  Second, we use the customized scoring matrix derived from
the estimated evolutionary model instead of the standard {\sc Blosum}
matrix.  {\sc Psi-blast} differs from our method in that it also uses
full length primary sequences and it effectively uses an empirical model
of position specific weight matrices to extract evolutionary information
from a set of multiple aligned sequences, without the benefit of using
a phylogeny and an explicit parametric model.

Compared to the {\sc Fasta} sequence alignment and {\sc Psi-blast} search,
our method can identify more alpha amylases.  In addition, because we
directly detect binding surface similarity instead of global sequence
similarity, our prediction has stronger implications for inferring
functional relationships. In contrast, {\sc Psi-blast} search does not
provide information about which residues are important for function.
We have also shown that our estimated rate matrix works much better than the
generic precomputed {\sc Jtt} matrix, especially when the query
template surface has a relatively small size.

\begin{table}[!h]
\captionsetup{labelsep=newline,singlelinecheck=false}
\begin{center}
\caption{\bf Detecting functionally related proteins.}
\vspace*{10pt}
\label{tab:dataset}
\begin{minipage}{\linewidth}
\renewcommand\thefootnote{\thempfootnote}
\begin{tabular}{lccc>{\bf}cccc}
\hline
\hline
Protein Family&Query &Pocket\footnote{Pocket id could be referenced through {\sc CastP} database ({\tt cast.engr.uic.edu}).}&pocket& Our\footnote{Our results are obtained from querying with a template binding surface and customize scoring matrices.} &Results by & Results by & ESD\footnote{The true answers are taken as those recorded in the human curated {\sc Esd} database.}\\
{}&{structure}&id&length&result&{\sc Psi-blast}\footnote{Results using {\sc Psi-blast} sequence alignment.}&{\sc Jtt}\footnote{Results using our method with a standard {\sc Jtt} matrix.}&(true answers)
\\
\hline
EC 3.2.1.1	&1bag &60  &18 &58 &45 &52 &75\\
EC 3.2.1.1	&1bg9 &61  &12 &48 &21 &8  &75 \\
\hline
EC 3.8.1.2 	&1qh9 &23  &16 &8   &8   &3   &8\\
EC 3.5.4.4 	&2ada &49  &28 &23  &17  &19  &23\\
EC 4.2.1.11	&1ebh &122 &35 &22  &20  &19  &22\\
EC 1.13.11.39	&1kw9 &34  &23 &18  &16  &18  &18\\
\hline \hline
\end{tabular}
\end{minipage}
\end{center}
\end{table}

To examine whether our method works for proteins of other
functions, we repeated our test using four additional enzymes of
different biochemical functions.  These are: 2,3-dihydroxybiphenyl
dioxygenase (E.C. 1.13.11.39), adenosine deaminase (E.C. 3.5.4.4),
2-haloacid dehalogenase (E.C. 3.8.1.2), and phosphopyruvate
hydratase (E.C. 4.2.1.11). As shown in Table~\ref{tab:dataset}, we
are able to find all other protein structures of the same E.C.\
numbers contained in the ESD in all four cases.  Our results
are better than using {\sc Psi-blast} or using the {\sc Jtt} matrix.

\section*{Discussion}
We have developed a Bayesian method for estimating residue
substitution rates. Bayesian inference of phylogeny was independently
introduced by Yang and Rannala \citeyearpar{Yang_MBE97}, Mau {\it et al} 
\citeyearpar{Mau_B99}, and Li {\it et al} \citeyearpar{Li_JASA00}.  Bayesian
methods have found wide applications
\citep{Huelsenbeck_S01,Huelsenbeck_SB02b}, including host-parasite
co-speciation \citep*{Huelsenbeck_E97}, estimation of divergence times
of species \citep*{Thorne_MBE98} , simultaneous sequence alignment
and phylogeny estimation \citep*{Mitchison_JME99}, inference of
ancestral states \citep*{Huelsenbeck_SB01}, and  determination of the root
position of a phylogenetic tree \citep*{Huelsenbeck_SB02a}.  Similar
to others, our approach is based on the Markov chain Monte Carlo
sampling technique.
Although we are not aware of any other studies using Bayesian models
for the direct estimation of substitution rates between amino acid
residues, our approach is a natural extension of existing work on 
maximum likelihood estimation \citep*{Goldman_MBE94,Yang_MBE98} of
codon substitution rates for amino acid residues, and other studies
based on Bayesian statistical analysis
(\citealt*{Yang_MBE97,Huelsenbeck_E97,Thorne_MBE98,Mau_B99}; \citealt{Huelsenbeck_S01}).

In this work, we studied the substitution of residues using amino acid
sequences instead of nucleotide sequences.  In our model, the
parameters of the continuous time Markov process are the rates of
direct substitutions between residues.  A more established model of
residue substitution is that of the substitutions between codons. This
model can provide rich information about detailed mechanisms of
molecular evolution.  For example, the differential effects of
transition {\it vs}.\ transversion and synonymous {\it vs}.\
nonsynonymous substitutions all can be modeled
\citep*{Goldman_MBE94,Yang_MBE98}.  Our choice of the current model of
direct residue substitution is based on two practical
considerations. First, for the application of predicting protein
functions, we find it is far easier to gather amino acid residue
sequences than nucleotide sequences when large scale database searches
are carried out. Second, when using scoring matrices derived from
substitution rates to detect remotely related proteins, amino acid
sequences give far better results in sensitivity and specificity
than nucleotide sequences \citep*{Pearson_JMB98,LioGoldman99_MBE}.
An interesting future study would be one that is based on codon
substitution models, which will help to identify possible bias in the 
current approach, where the effects of transition/transversion and
synonymous/nonsynonymous substitutions are not considered.

It has long been recognized that the evolutionary divergence of protein
structures is far slower than that of sequences \citep*{Chothia_EJ86}.
Since physical constraints on protein structure would give rise to
associations between patterns of amino acid replacement and protein
structure \citep*{Koshi_PSB96,Koshi_P97}, the substitution rates of
residues in different secondary structural environments and of
different solvent accessibility have been well-studied
\citep*{Lesk_JMB82,Goldman_JMB96,Goldman_G98,Thorne_MBE96,Bustamante_MBE00}.
In a pioneering work, Thorne {\it et al}.\ developed an evolutionary
model that combines secondary structure with residue replacement, and
showed that the incorporation of secondary structure significantly
improves the evolutionary model for sucrose synthase
\citep*{Thorne_MBE96}.  The impact of secondary structure and solvent
accessibility on protein evolution was further studied in detail using
a hidden Markov model in \citep*{Goldman_G98}.  Additional work showed
that an accurate evolution model can in turn lead to accurate
prediction of protein secondary structure
(\citealt*{Goldman_JMB96}; \citealt{Lio_B98}).  Parisi and Echave have further
developed a simulation model to study the effects of selection of
structural perturbation on the site-dependent substitution rates of
residues (\citealt*{Echave01_MBE}; \citealt{Robinson_MBE03}; \citealt*{Parisi_G05}).  These studies
highlighted the importance of physical constraints 
on protein evolution.

Our work is a continuation in the direction of assessing substitution
rates of residues in different structural environments, but with an
important novel development.  Here we proposed to study substitution
rates of residues in a new structural category, namely, residues from
local binding surface regions that are directly implicated in
biochemical functions.  Since a fundamental goal of studying protein
evolution is to understand how biological functions emerge, evolve,
and disappear (\citealt*{Gu_G03}; \citealt{Vogel_COSB04}; \citealt*{Lecomte_COSB05}), estimation
of the substitution rates of residues on functional surfaces is
critically important.

Proteins are selected to fold to carry out necessary cellular roles.
In many cases, they are involved in binding interactions with other
molecules.  Surface binding pockets and voids are therefore the most
relevant structural regions, which can be computed using exact
algorithms \citep*{Liang98_PS}.  A unique
advantage of this novel structural category is that it allows better
separation of residues experiencing selection pressure due to the
constraints of biochemical functions from those due to the constraints
for physical structural integrity.  In contrast, the
structural categories of residues in different secondary structural
environments and solvent accessibility are more suited to study how
substitutions are related to protein stability, because they
inevitably will include many conservation patterns due to the
requirements of structural stability.  

For example, solvent accessibility directly relates to the driving
force of hydrophobic effects for protein folding, and secondary
structures are essential for maintaining protein stability
 (\citealt*{Dill_B90}; \citealt{Dill_PS95}). The structural categorizations developed
in \citep*{Goldman_JMB96,Goldman_G98,Thorne_MBE96} are well-suited for
studying how protein evolution is constrained by physical interactions
important for protein folding and stability.  For example, the
patterns of hydrophobic residues in the buried interior, polar residues on
the surface, and small residues in $\beta$-turns are all due to structural
constraints and do not have direct functional implications.  Indeed,
the study of Koshi and Goldstein found strong correlation between
transfer free energy $\Delta G$ of amino acid residues, a
physico-chemical property of amino acid solvation energy, and residue
substitution rates \citep*{Koshi_PSB96}.  The categorization of residues
proposed here are designed for studying how protein evolution is
constrained by function ({\it i.e.}, protein-ligand/substrate binding
and protein-protein interactions).  To our best knowledge, this is the
first study in which a structure-derived category amenable for
computation is proposed that separates residues selected for function
from residues selected for stability.

Our results  showed that residues located in functional pockets
have different substitution rates from residues in the remaining
parts of the protein.  The differences are mostly due to residues such
as His and Asp that are known to be important for protein
function.  All of these region-specific substitution rate matrices
are different from the precomputed {\sc Blosum} matrix.

It is informative to examine the difference of the substitution
rates in the {\sc Jtt} matrix and the binding site specific rate
matrices we estimated.  The {\sc Jtt} matrix was developed using a
very large database of sequences, and the overall composition
$D_{\mbox{{\sc Jtt}}}$ of amino acid residues is very different
from the composition $D$ of the binding surfaces.  Hence, the
conserved residues, or the values of the diagonal elements
$s_{ii}$ of the substitution matrix, are very different. This is
reflected in the different residue composition for functional
surfaces and the full protein sequence
(Figure~\ref{fig:PocDist}b).  This would result in different
overall patterns of substitutions.  For substitution after a long
time interval, it is necessary to estimate the off-diagonal
elements $s_{ij}$ with some accuracy, as the substitutions would
accumulate with time, and identifying remotely related binding
surfaces becomes difficult.

It is challenging to estimate substitution rates of amino acid
residues in a local region.  The number of residue positions for a
specific region may be small, and the available sequences in the
phylogenetic tree may also be limited.  It is unlikely that all 189
independent substitution rates of the $20\times 20$ matrix can be
estimated accurately when only limited data is available.  In this
study, we can only estimate substitution rates for occurring pairs,
namely, substitutions between residues that occur in the same position
in different sequences.  However, for applications such as inferring
protein functions by matching similar binding surfaces, our results
show that the constructed scoring matrices are very effective.  It is
likely that the substitutions (or lack thereof) that occur in the
sampled data for a specific region are the most important ones in
overall patterns of evolution of residues in this specific region.
For example, the most important features in a functional pocket on a
protein structure are the conserved residues.  Accurate estimation of
the diagonal rates ($s_{ii}$) is therefore the most important task.
Because conserved residues appear in relatively higher frequency, they
often can be estimated well.  If some substitutions never occur in the
sampled data, they probably are not important and setting their values
to a baseline offset value such as that from a uniform prior would be
reasonable. We have carried out detailed studies on identifying
functionally related alpha amylases and other enzymes by querying with
one or more template binding surface and assessing similarity using
scoring matrices derived from the estimated rates.  As shown in
Table~\ref{tab:dataset}, our approach works very well in practice.  In
a control study, we assign random values to the matrix entries, which
conform to the normalization condition.  Scoring matrices derived from
this randomized rate matrix are ineffective, and we were not able to
find any functionally related proteins for any example listed in
Table~\ref{tab:dataset}.

One might wish to estimate a $20\times20$ substitution
rate matrix that is specific to an individual site or position in the
sequence.  However, this would require a very large amount of data
that are not available in practice.
In addition, it is conceivable that estimating site specific rate
matrices may not be necessary or possible.  For example, if a residue is critical
for protein folding stability, it might be conserved through all
stages of the evolution, and there is no variation at this particular
position of the amino acid sequences.  In such cases, it is difficult
to estimate a full substitution matrix for this site.  In our approach,
we essentially pool residues that are located in the same region
together, and assume they experience similar evolutionary
pressure.

Ultimately, the effectiveness of incorporating structural information
in phylogenetic analysis and evolutionary models can be tested on the
criterion whether it in turn helps to understand the organization
principles of protein structures and their biochemical functions.  
As
indicated by  successful applications in protein function
prediction reported here, structure-based phylogenetic analysis
provides a powerful framework for studying significant problems in
structural biology.

Our method benefits from existing computational techniques.  Without
the mathematical theory that formalizes our intuitive notion of protein
shapes such as pockets and voids \citep*{Edels98_ADM}, efficient
algorithms for their computation \citep*{Edels98_ADM,Liang98_PS},
strategies for shape similarity assessment \citep*{Binkowski03_JMB}, as
well as demonstrated success of these computational techniques
\citep*{Liang98_PS,LiLiang03_Proteins,Binkowski03_JMB,LiLiang05_Proteins}, the
novel category of functionally important surface pockets would not be
possible.  

There are, however, some limitations in our method. If the number of
homologous sequences is too few ($<10$) or the length of the
functionally important binding pocket is too short ($<8$ residues),
there will not be enough data for parameter estimation.  
Another limitation of our study is the assumption that all sites in a
protein evolve according to the same rate matrix along all branches of
the phylogenetic tree.  Although simulation studies and applications
indicate that the estimated rates are sufficiently accurate for the
purpose of detecting functionally related protein surfaces, this
assumption may not be realistic for studying detailed evolutionary
history and mechanisms for a specific protein
\citep*{Yang_MBE93,Yang_JME94b,Huelsenbeck_JME99,Felsenstein_JME01}.

Our simulation study is simple and cannot provide a full picture of
the estimation errors under different biological conditions.  The
focus of our simulation study is to assess how estimation error is
affected by the length of a functional pocket.  
In our method, the proper and accurate construction of a high quality
phylogenetic tree is essential.  We find it important to carefully
select amino acid sequences to ensure quality multiple sequence
alignments, where few gaps are introduced and proteins of different
divergence are well represented.  In our practice, we find that the
maximum likelihood estimator of {\sc Molphy} works well with amino acid
sequences for constructing phylogenetic trees. 
The effects of the assumption that the input phylogenetic tree is
optimal, as well as the effects of different input branch lengths
on the accuracy of estimation, needs further detailed studies.  Our
preliminary results suggest that the estimated scoring matrices for
protein functional sites and database search results are insensitive to
small perturbations in the phylogenetic tree and the branch
lengths.  For instance, in a database search of alpha amylase, we are
able to use different surface templates, each from a different protein
structure with its own slightly different phylogenetic tree and branch
lengths.  Our results show that the sets of functionally related
proteins are nearly identical (data not shown).

Furthermore, the choice of a prior is an important and complex issue
in Bayesian statistics.  We assume that the likelihood function
dominates and the information from the prior is limited.  More
detailed study is needed for a clear understanding of the influence of
the choice of prior.

In summary, we have extended existing continuous time Markov models
of residue substitution from that of codon-codon replacement to a
model of residue-residue replacement.  We have also developed a novel
structural category of local surface regions that is well-suited for
studying the evolution of protein functions.  We have implemented an
effective Bayesian Monte Carlo method that can successfully estimate
the substitution rates of residues in small local structural regions in 
proteins.  In addition, we have developed a database search method
using scoring matrices derived from estimated residue substitution
rates.  Our results in solving the fundamental problem of inferring
protein functions from protein structures show very encouraging
results.  There are other novel technical developments.  For example,
we find it necessary to develop an efficient move set for rapid mixing
in Monte Carlo estimation of substitution rates.  We have also explored
how reliability of estimated substitution rates depends on the size of
the local region.  As indicated by the successful applications
reported here, we believe that phylogenetic
analysis of protein evolution provides powerful tools for the
important bioinformatic task of protein function prediction.

\section*{Acknowledgment}
We thank Dr.\ Andrew Binkowski for help in {\sc pvSoar} search,
Ronald Jackups, Jr for proofreading the manuscript, Drs.\ Rong Chen,
Susan Holmes, Art Owen, and Simon Whelan for rewarding discussions.
We also thank Dr.\ Jeffrey Thorne and an anonymous referee for
insightful suggestions.  This work is supported by grants from NSF
(CAREER DBI0133856), NIH (GM68958), and ONR (N000140310329).

\renewcommand\refname{Literature Cited}
\bibliography{MCMC,ref}

\bibliographystyle{mbe}

\end{document}